\documentclass[preprint2]{aastex62}
\usepackage{graphicx}
\usepackage{color}
\usepackage{mathrsfs}

\newcommand{\km}{\,{\rm km}} 
\newcommand{\erg}{\,{\rm erg}} \newcommand{\yr}{\,{\rm yr}}
\newcommand{\ps}{\,{\rm s}^{-1}}

\newcommand{\pc}{\,{\rm pc}}

\newcommand{\pcc}{\,{\rm cm}^{-3}}
\newcommand{\psc}{\,{\rm cm}^{-2}}

\newcommand{\snr}{G106.3+2.7}
\newcommand{\NHH}{N({\rm H}_2)}
\newcommand{\HH}{({\rm H}_2)}
\newcommand{\twCO}{$^{12}$CO}
\newcommand{\thCO}{$^{13}$CO}

\newcommand{\otz}{$J$=\,1--0}
\newcommand{\tto}{$J$=\,2--1}

\newcommand{\du}{d_{0.8}}

\newcommand{\h}{$^{\rm h}$}
\newcommand{\m}{$^{\rm m}$}
\newcommand{\s}{$^{\rm s}$}
\newcommand{\ci}{$^\circ$}

\begin{document}

\begin{tiny}

\end{tiny}

\title{IRAM 30\,m CO-line Observation toward PeVatron Candidate \snr: 
Direct Interaction between the Shock and the Molecular Cloud 
Remains Uncertain}
\author{Qian-Cheng Liu$^1$, Ping Zhou$^{1,2}$, Yang Chen}
\affil{Department of Astronomy, Nanjing University, 163 Xianlin Avenue, Nanjing 210023, People's Republic of China; ygchen@nju.edu.cn, pingzhou@nju.edu.cn}
\affil{Key Laboratory of Modern Astronomy and Astrophysics, Nanjing University, Ministry of Education, Nanjing 210093, People's Republic of China}

\begin{abstract}
Supernova remnant (SNR) \snr\ was recently found to be
one of the few potential Galactic hadronic PeVatrons.
Aiming to test how solid the SNR is associated with the 
molecular clouds (MCs) that are thought to be responsible
for hadronic interaction,
we performed a new CO observation with the IRAM 30\,m telescope toward
its ``belly'' region,
which is coincident with the centroid of the $\gamma$-ray emission. 
There is a filament structure in the local-standard-of-rest
velocity interval $-8$ to 
$-5\km\ps$ that nicely follows the northern radio boundary of the SNR.
We have seen asymmetric broad profiles of \twCO\ lines, with widths
of a few $\km\ps$, along the northern boundary and in the ``belly'' region
of \snr, but similar \twCO\ line profiles are also found
outside the SNR boundary.
Further, the low \twCO\ \tto/\otz\ line ratios 
suggest the MCs are cool.
Therefore, it is still uncertain whether the 
MCs are directly disturbed by the SNR shocks,
but we do find some clues that the MCs are nearby and
thus can still be illuminated by the escaped protons from the SNR.
Notably, we find an expanding molecular structure with a velocity
of $\sim 3.5\km\ps$ and a velocity gradient of the MCs across the SNR
from $\sim -3$ to $-7\km\ps$,
which could be explained as the effect of the wind blown
by the SNR's progenitor star.
\end{abstract}

\keywords{ISM: individual objects (\snr) -- ISM: supernova remnants -- ISM: molecular clouds}

\section{Introduction} \label{introduction}
Supernova remnants (SNRs) are suggested to be one of the major factories to produce
cosmic rays (CRs) up to $\sim 10^{15}$\,eV. High-energy GeV/TeV $\gamma$-ray
photons can be produced via either the interaction between the CR protons and the 
gas material (hadronic origin) or inverse-Compton scattering of relativistic electrons 
(leptonic origin). 
The former scenario requires a dense environmental medium, such as
molecular clouds (MCs), while the latter origin is not subject to the 
nearby density.
Therefore, an investigation of the molecular
environment of an 
$\gamma$-ray-emitting SNR is vital to distinguish the origin of the 
high-energy emission.

\begin{figure*}[t]
\centering
\includegraphics[scale=.4]{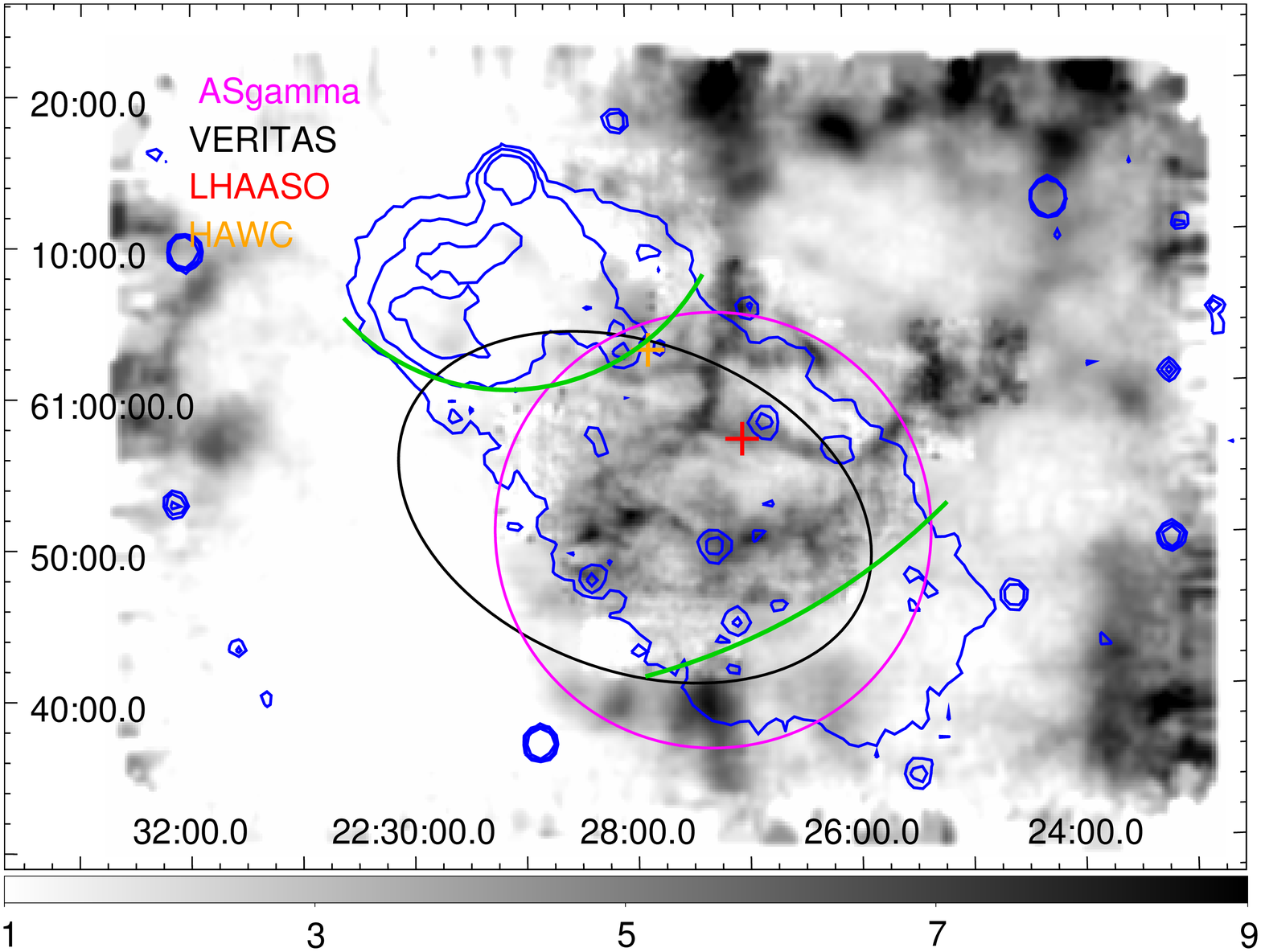}
\caption{\twCO\ \otz\ intensity map integrated in velocity range
$-10$ to $-2\km\ps$, overlaid by 1.4\,GHz radio contours from the 
CGPS \citep{tay03}
in levels 6.9, 7.3, and 8\,K.
The magenta circle and black ellipse represent the 
gamma-ray emission detected by the Tibet AS$\gamma$ \citep{ame21}
and VERITAS \citep{acc09}, respectively.
The red and orange crosses show the best-fit positions of the LHAASO
\citep{cao21} and HAWC \citep{alb20} sources.
The green arcs crudely separate the ``head'', ``belly'',
and ``tail'' regions (referring to the upper, middle, and lower parts, 
respectively) of the SNR described in this paper.
The upper arc is adopted from \cite{ge20}, and the lower arc
follows the southwestern edge of the field of view of our
IRAM observation.
}
\label{multi}
\end{figure*}

\begin{figure*}
\centering
\includegraphics[scale=.5]{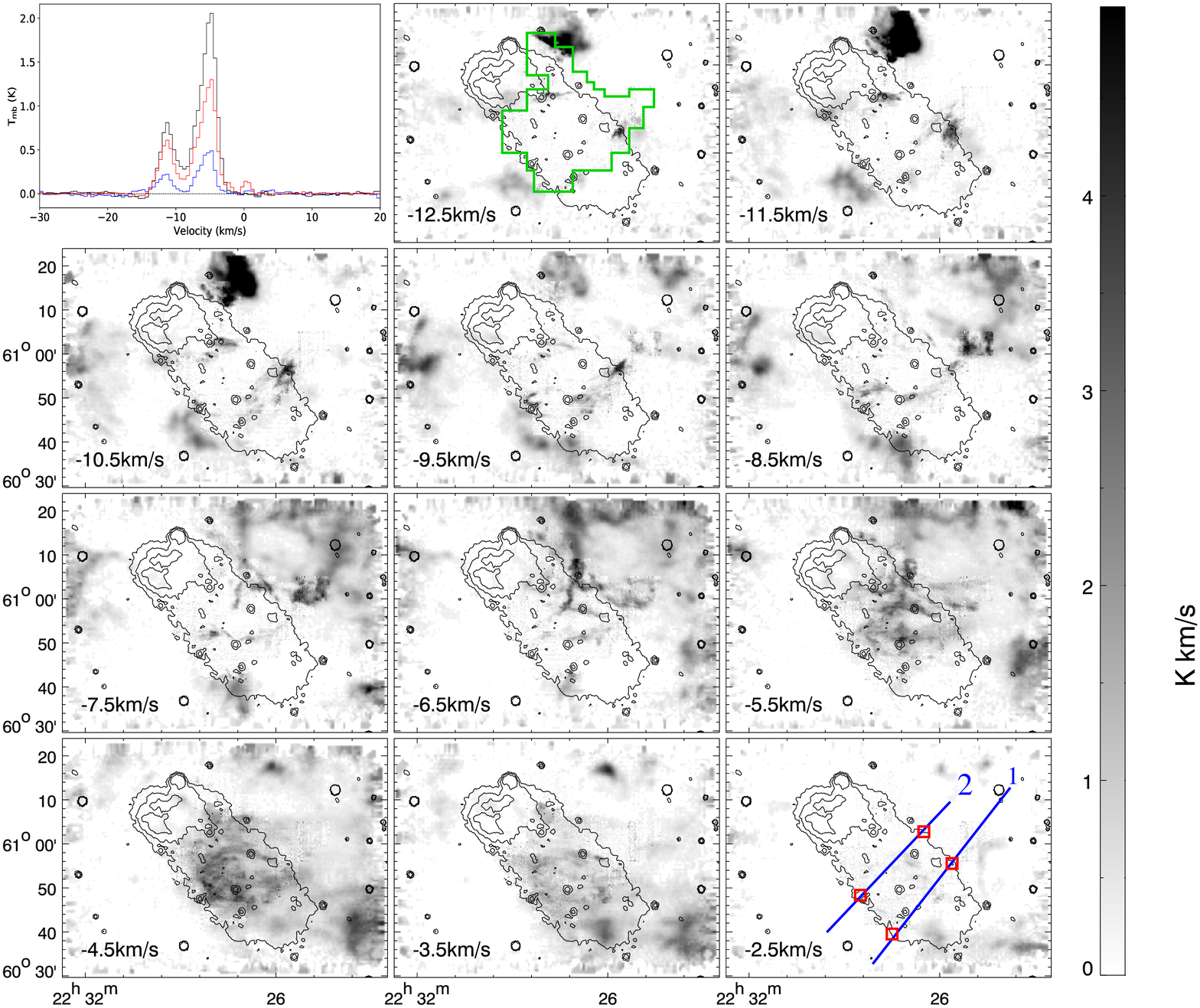}
\caption{ \twCO\ \otz\ intensity map integrated at $1\km\ps$ in the velocity 
range $-13$ to $-2\km\ps$, overlaid by 
radio contours 
the same as those in Figure~\ref{multi}.
The region enclosed with the green polygon in the top middle panel delineates the 
region we observed with the IRAM 30\,m telescope,
and the data within this region for all the panels are from IRAM 30\,m
telescope.
The data outside this region are from the archived PMOD.
The data from the IRAM 30\,m telescope have been resampled and
re-grid to match the velocity resolution and pixel size of the PMOD data, 
and
the data from PMOD has been multiplied by a factor of 2 for better visibility. 
The top left panel shows the averaged CO spectra of the polygon region in 
the velocity range $-30$ to $20\km\ps$ from the IRAM observation. 
The black line is for \twCO\ \otz, the red line is for \twCO\ \tto, the blue 
line is for \thCO\ \otz, and the dashed line is for the 0\,K main-beam 
temperature. The \thCO\ \otz\ line has been multiplied by a factor of
three for better visibility.
The lines used to make the position-velocity diagrams in Figure~\ref{pv} are 
indicated in the bottom right panel in dark blue.
}
\label{chanelmap}
\end{figure*}

SNR \snr\ has attracted substantial attention recently, 
as it was found to be one of the
few potential Galactic hadronic PeVatrons \citep{alb20,ame21,cao21}.
Extended $\gamma$-ray emission has
been detected with Fermi-LAT \citep{xin19}, Milagro \citep{abd07}, 
VERITAS \citep{acc09}, 
and the Tibet AS$\gamma$ experiment \citep{ame21}
from \snr\ (see Figure~\ref{multi}).
Recently, nonthermal X-ray emission was also discovered 
based on the analysis of \emph{Chandra}, \emph{XMM-Newton} and 
\emph{Suzaku} data \citep{ge20,yut21}. 
There is a bright X-ray point source 
toward the radio peak of \snr\ \citep{hal01b}, 
and it was later confirmed to be a pulsar 
in the radio, X-rays \citep{hal01}, and the
$\gamma$-ray \citep{abd09} bands.
The existence of a pulsar establishes the core-collapse origin of \snr.

\snr\ has a comet-shaped morphology in the radio band, with a bright head 
in the northeast and a tail toward the southwest \citep{pin00}.
It was suggested to be located at the edge of a large HI bubble and 
associated with the MCs at a systemic 
local-standard-of-rest (LSR) velocity of about 
$-6\km\ps$ based mainly on the spatial correspondence \citep{kot01}. 
This HI bubble was suggested to be created by stellar winds and 
SNe in the previous generation.
Furthermore, the head of the SNR is found to be located in a
small HI bubble \citep{kot01}.
Spatial coincidence has also been used to indicate a possible 
association between the $\gamma$-ray emission, the SNR, and the MCs
\citep[e.g.,][]{acc09, xin19, ame21, bao21}. 
However, the nature of the $\gamma$-ray emission (hadronic or leptonic
origin) is still under debate 
\citep[e.g.,][]{acc09, xin19, alb20, lius20},
and there is little dedicated investigation 
that aims at kinematically clarifying if the 
MCs are directly disturbed by the SNR shock (e.g., asymmetric 
broad-line profiles of \twCO\ lines, 
high \twCO\ \tto/\otz\ line ratios, etc.;
\citealt{jia10, che14}).
A new, independent CO observation with higher spatial and velocity 
resolution is therefore needed, which helps explore
the astrophysical environment toward the direction of \snr.

Here, we present a new CO-line observation using the IRAM 30\,m telescope 
toward \snr\ and search for kinematic and physical signatures of the SNR-MC
interaction.
The observation and the data reduction process are described in 
Section~\ref{observation}, and the results are presented in 
Section~\ref{results}; the discussion of the main results
and the summary are presented in Section~\ref{discussion} and \ref{summary},
respectively.

\begin{figure*}
\centering
\includegraphics[scale=.5]{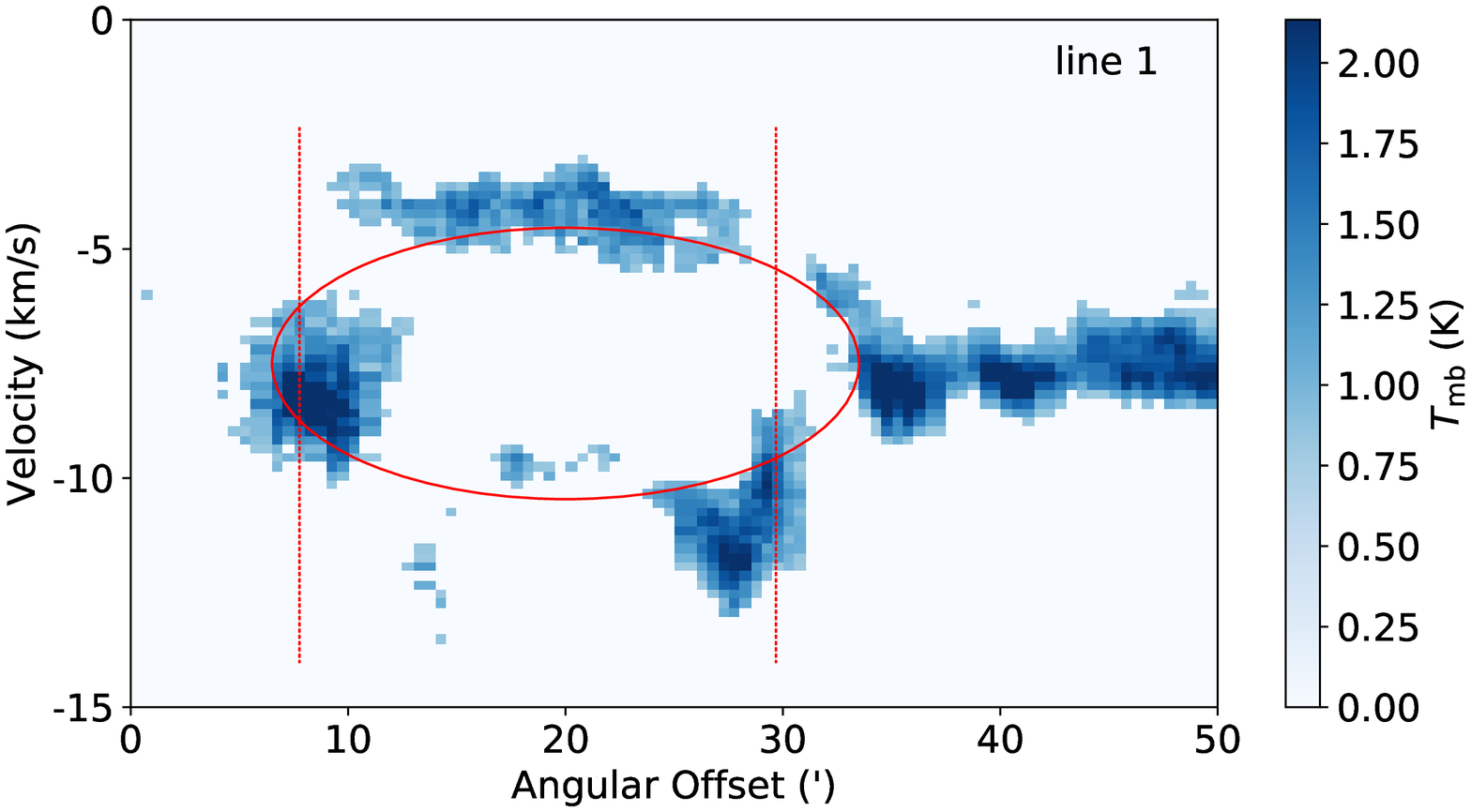}
\includegraphics[scale=.5]{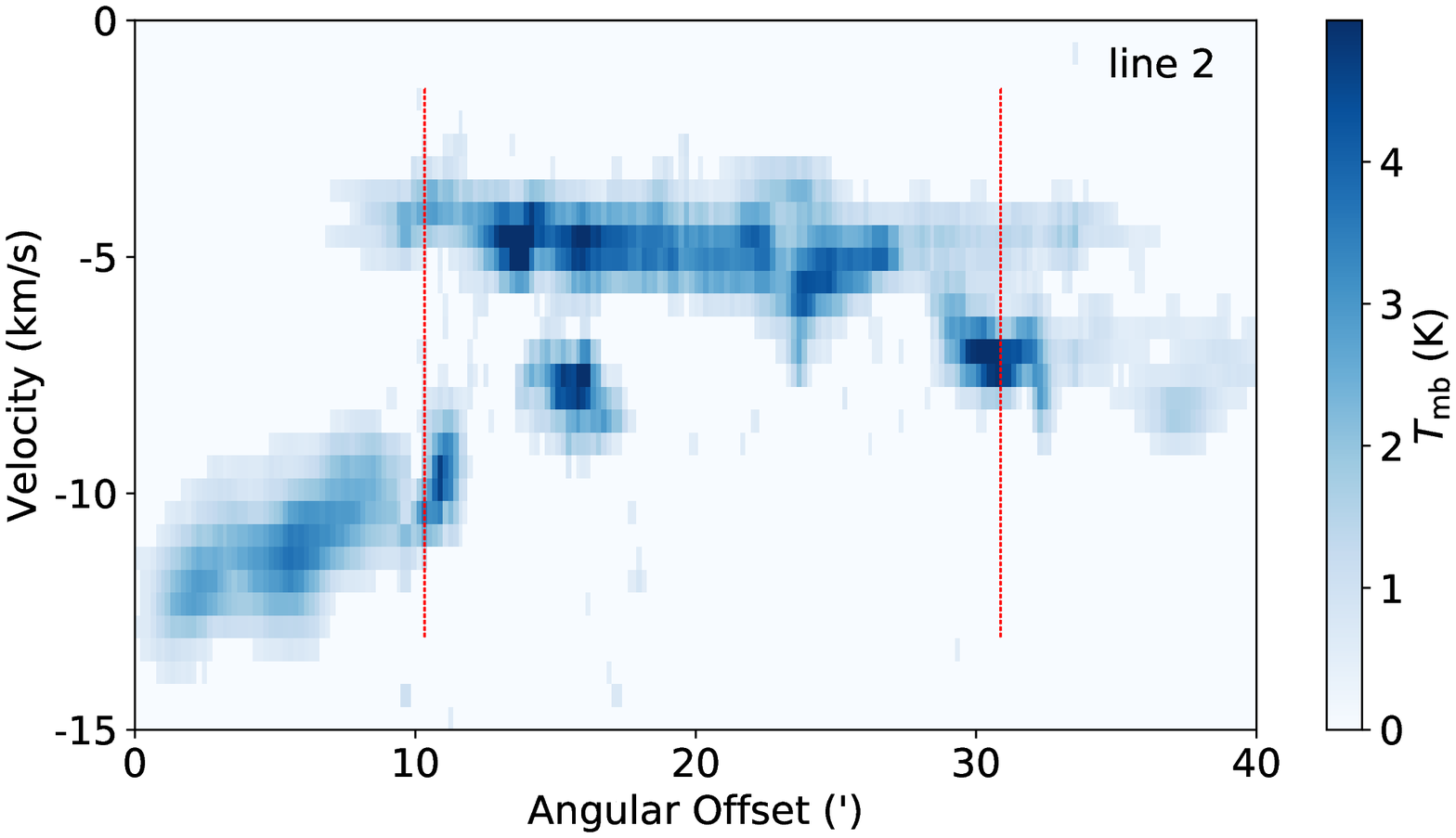}
\caption{Position-velocity diagrams of \twCO\ \otz\ emission for two lines
that go 
from R.A.=22\h28\m07\s.58, Decl.=$+60$\ci33$'16.92''$ to
R.A.=22\h23\m54\s.65, Decl.=$+61$\ci12$'39.6''$ with a 
$52.4^\circ$ inclination
(line1, the top panel), and 
from R.A.= 22\h29\m31\s.56, Decl.=$+$60\ci40$'46.92''$ to 
R.A.=22\h25\m45\s.84, Decl.=$+61$\ci09$'54.00''$ 
with a $46.7^\circ$ inclination
(line2, the bottom panel), respectively (see the bottom right panel of 
Figure~\ref{chanelmap}). The red lines in each diagram delineate the 
edge of \snr\ (see the red small rectangles in the bottom right panel of
Figure~\ref{chanelmap}), and the ellipse in the top panel indicates the 
expanding molecular structure in the position-velocity diagram.
}
\label{pv}
\end{figure*}

\section{Observation and data reduction} \label{observation}
Our observations of molecular lines toward \snr\ was made simultaneously 
in \twCO\ \otz, \twCO\ \tto, and \thCO\ \otz\ with the IRAM 30\,m telescope 
during 2020 December, 22 -- 30, for a total of 40\,hr. 
The observations covered an irregular polygon region toward the ``belly'' 
region of \snr\ with an area of about 680\,arcmin$^2$ (see the top 
middle panel of Figure~\ref{chanelmap}). 
The mapping was conducted with the on-the-fly position-switching mode, 
using the Eight Mixer Receiver in E0 and E1 band and the fast Fourier 
transform spectrometers (FTS).
The backend FTS provided a bandwidth of 16\,GHz and a spectral resolution 
of 200\,kHz. 
The velocity resolution is thus $\sim 0.5\km\ps$ for \twCO\ \otz\ 
and \thCO\ \otz, and $\sim 0.25\km\ps$ for \twCO\ \tto.
The half-power beamwidth (HPBW) of the telescope is about $21''$ at 
115\,GHz and about $ 10''.7$ at 230\,GHz, and the main-beam 
efficiency is about 0.78 at 115\,GHz and about 0.59 at 230\,GHz.
To better study the three CO transitions, we have convolved the 
angular resolution to $22''$ and resampled the velocity resolution 
to $0.5\km\ps$ for all line data. 
The rms noise levels of the main-beam temperature ($T_{\rm mb}$) of the 
convolved and resampled data are
$\sim 0.07$ -- 0.2\,K for \thCO\ \otz\ lines, $\sim 0.2$ -- 0.8\,K 
for \twCO\ \otz\ lines, and 
$\sim 0.06$ -- 0.2\,K for \twCO\ \tto\ lines.
All the IRAM 30\,m data were reduced with 
the GILDAS/CLASS package developed by the IRAM
observatory \footnote{http://www.iram.fr/IRAMFR/GILDAS}.

To study the larger-scale environment of the SNR, we have used the archived 
\twCO\ \otz\ data obtained with the 13.7\,m millimeter-wavelength 
telescope of the Purple Mountain Observatory at Delingha (PMOD). 
The HPBW of the data is about $50''$, and the velocity resolution of the data 
is about $0.16\km\ps$. 
The rms noise level of the data varies in a range of
$\sim 0.1$ -- 0.2\,K. 
We also used 1.4\,GHz radio continuum emission data
from the Canadian Galactic Plane Survey \citep[CGPS;][]{tay03}.

\begin{figure*}
\centering
\includegraphics[scale=.7]{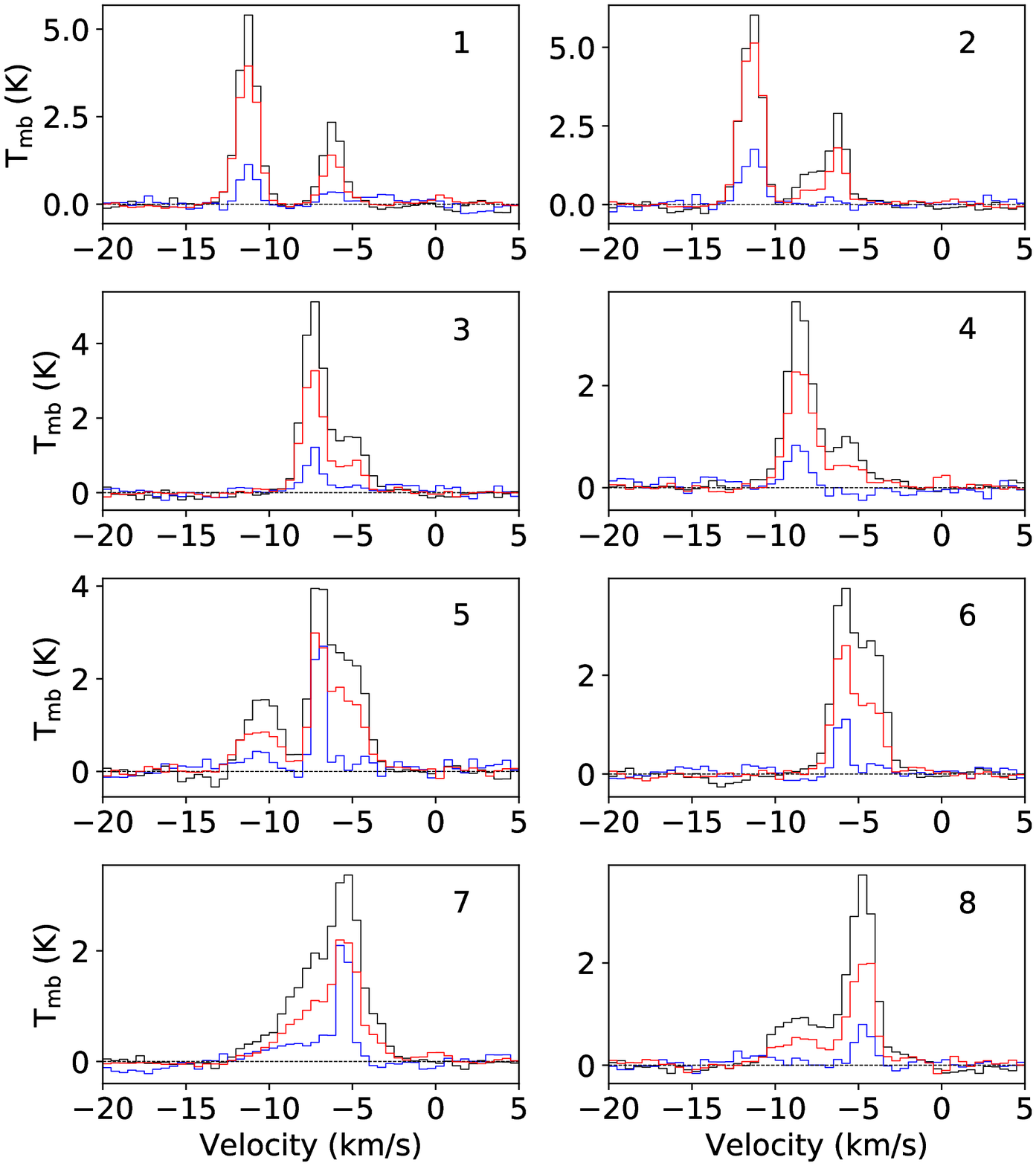}
\caption{ Averaged spectra of regions ``1'' to ``8'' in the velocity range $-20$ to
$5\km\ps$. The eight regions are defined in Figure~\ref{tkin}. The black lines
denote the \twCO\ \otz\ spectra, the red lines denote the \twCO\ \tto\ spectra,
the blue lines denote the \thCO\ \otz\ spectra, and the dashed lines represent
the 0\,K main-beam temperature. The \thCO\ \otz\ spectra have been multiplied by
a factor of three for better visibility.
}
\label{regsed}
\end{figure*}

\section{Results} \label{results}

\subsection{Spatial Distribution of the Molecular Clouds}\label{spatial}
The top left panel of Figure~\ref{chanelmap} shows the 
\twCO\ \otz, \twCO\ \tto, and \thCO\ \otz\ spectra averaged
from the polygon region that was observed with the IRAM 30\,m telescope. 
There are two prominent CO emission peaks, at around $-11\km\ps$ 
and $-4.5\km\ps$.
To examine the spatial distribution of these MCs, we have made \twCO\ \otz\ 
emission channel maps in the velocity range $-13\km\ps$
to $-2\km\ps$ using the highest spatial resolution CO data to date
(Figure~\ref{chanelmap}, where we have inserted the data from the 
IRAM 30\,m telescope into the data from the PMOD). 
The maps cover the main components of the two emission peaks. 

There are a few noticeable
structures in this velocity range.
(1) In velocity interval $-8$ to $-5\km\ps$,
except for the envelope around the head region 
of the SNR as noticed by \cite{kot01},
there is a filament-like structure nicely aligning the 
northwestern edge 
of the SNR, with an orientation from the northeast to the southwest.
(2) In velocity interval $-6$ to $-3\km\ps$,
there is a ring-like structure and some diffuse \twCO\ emission
toward the ``belly'' region 
(see Figure~\ref{multi})
of the SNR. 
Notably, the spatial distribution of the diffuse \twCO\
emission happens to overlap with 
the $\gamma$-ray emissions detected by Fermi-LAT 
\citep{xin19} and the Tibet AS$\gamma$ experiment \citep{ame21},
while the head/PWN part of the SNR seems faint in \twCO\ line.
Actually, the LSR velocity of the MCs toward the ``belly'' region is 
very similar to that of environmental HI gas \citep{kot01}.

\subsection{Position-Velocity Diagram of the Molecular Clouds} 
\label{pvresult}
We note some molecular structures with unusual velocity features 
toward the ``belly'' region of the SNR.
Figure~\ref{pv} shows the 
position-velocity (P-V) diagrams of slices taken along the two lines,
labeled in the bottom right panel of Figure~\ref{chanelmap}.
The P-V diagram along line 1 shows an expanding molecular structure,
with an angular extent larger than that of the SNR.
This expanding structure is centered at
$\sim -7.5\km\ps$ with an expansion velocity of $\sim 3.5\km\ps$
and an angular size of $\sim 26'$.
Also, we find a velocity gradient that changes from $\sim -3\km\ps$ at
angular offset $10'$ to $\sim -7\km\ps$ at angular offset $30'$
along line 2.
The measured velocity variation is about
$4\km\ps$, which is similar to the expansion velocity of the expanding
structure.
Because there lacks conclusive kinematic evidence 
of disturbation of the MCs by the shock of the SNR
(see \S~\ref{line} and \ref{lineratio}), the P-V structures
could be a result of other processes, such as stellar wind(s), 
instead of the shock of SNR \snr.

\begin{figure*}
\centering
\includegraphics[scale=.9]{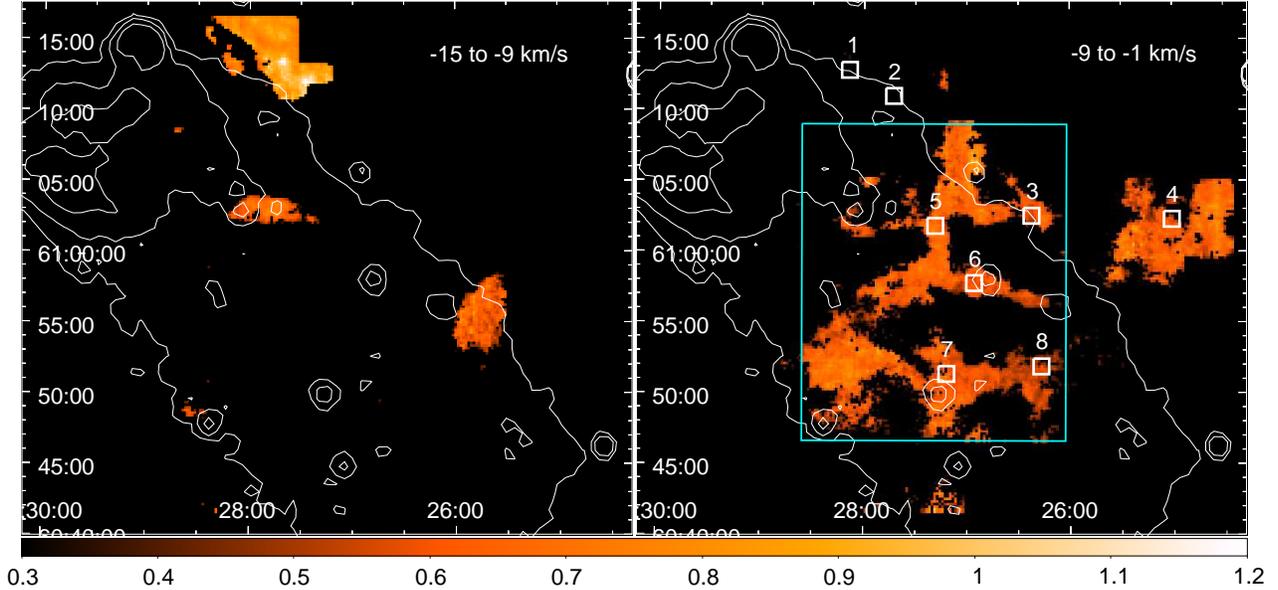}
\caption{ Ratio maps of IRAM \twCO\ \tto/\otz\ for LSR velocity ranges
$-15$ to $-9\km\ps$ (left panel) and $-9$ to $-1\km\ps$. The pixels toward
which the value are smaller than $5\sigma$ for either \twCO\ \tto\ or 
\twCO\ \otz\ have been omitted. 
The regions delineated by white rectangles are used to 
extract CO spectra (Figure~\ref{regsed}), and the region delineated by the cyan
rectangle is used to estimate the properties of the MC toward the ``belly''
region of \snr\
in velocity range $-9$ to $-1\km\ps$ (Section~\ref{discussion}).
The radio contours are the same as those in Figure~\ref{multi}.
}
\label{ratio}
\end{figure*}

\subsection{Molecular Line Profiles}\label{line}
The physical properties of an MC can be strongly affected if it is 
shocked by the SNR blast wave.
Observationally, the \twCO\ line profiles can be one/two-sided broadened 
if the MC is located at the edge of or within the SNR, and
such kinematic signatures are present in many SNR-MC interacting systems
like IC\,443 \citep[e.g.,][]{whi87}, W28 \citep[e.g.,][]{ari99},
Kes\,75 \citep{su09}, 3C\,397 \citep{jia10}, Kes\,78 \citep{zho11},
G357.7+0.3 \citep{rho17}, and CTB\,87 \citep{liu18}.

To examine whether 
SNR \snr\ is interacting with the MCs,
we have inspected the CO line profiles toward the region observed by 
the IRAM 30\,m telescope
(see Appendix~\ref{linegrid}).
Figure~\ref{regsed} shows the averaged emission line profiles delineated 
by the white rectangles in Figure~\ref{tkin}, where regions 1 to 3 are 
at the edge of the SNR, region 4 is outside the SNR boundary,
and regions 5 to 8 are in the ``belly'' 
region, projectively.
There are separated \twCO\ components at $\sim -11$ and $\sim -6\km\ps$ 
in both regions 1 and 2. 
The \twCO\ components at $\sim -6\km\ps$ are weaker and show 
weaker \thCO\ line emission, which is either the red-wing 
broadened part of the main components at $\sim -11\km\ps$
or a separated, irrelevant velocity component.
The \twCO\ line profiles in regions 3 to 6 are similar to those
in regions 1 and 2, except that the main components are at $\sim -8\km\ps$
in regions 3 to 5 and at $\sim -6\km\ps$ in region 6.
Further, there is a weak peak at $\sim -11\km\ps$ in region 5.
There are main \twCO\ components at $\sim -5\km\ps$ in both
regions 7 and 8, while there seem to be blue-wing broadened parts at
$\sim -11$ to $-7\km\ps$. 
We note that the profiles of the \twCO\ lines are systematically 
asymmetric 
in the ``belly'' region of the SNR (regions 1--3 and 5--8). 
However, 
there are similar \twCO\ line profiles in the region 
outside the SNR boundary (region 4).
Furthermore, the widths of the asymmetric wings are only a few $\km\ps$,
indicating that the disturbance in the molecular gas
is not strong.
Therefore, it is uncertain whether 
the broadened profiles in the region projected inside the SNR
are the results of interaction between the SNR shock and the MC.
Alternatively, they could be a signature
of other perturbed processes, such as 
perturbation by
the stellar wind(s).

\subsection{The \twCO\ \tto/\otz\ Line Ratio} \label{lineratio}
The \twCO\ \tto/\otz\ line ratios 
($R_{21/10}$) at the CO line wings could be enhanced 
($\ga 1$) if the molecular gas is disturbed and heated by the 
SNR shock 
\citep[e.g.,][]{set98, jia10, san19, san20}.
By inspecting the spectral properties of the CO emissions,
we have not found enhanced \twCO\ \tto/\otz\ line ratios 
at the CO line wings toward the region observed with IRAM 30\,m 
telescope (see Figures~\ref{regsed}, 
\ref{linegrid_N}, \ref{linegrid_S}, and \ref{linegrid_W}).

Figure~\ref{ratio} shows the spatial distribution
of the ratio $R_{21/10}$ of the molecular gas in velocity 
intervals $-15$ to 
$-9\km\ps$ and $-9$ to $-1\km\ps$.
We do not show the pixels where the values are smaller 
than $5\sigma$ for either
\twCO\ \tto\ or \twCO\ \otz\ emission.
Although the \twCO\ \tto/\otz\ line ratios in some pixels
just outside the northern edge of the SNR could reach unity in velocity 
interval $-15$ to $-9\km\ps$,
they mainly arise from the CO line centers instead of the line wings
(see Figure~\ref{linegrid_N}).
This means that this MC with high $R_{21/10}$ has a relatively
higher kinetic temperature (see Appendix~\ref{temperature})
but is not shocked and is a quiescent
MC outside the SNR.
On the other hand, the line ratios in all other pixels are clearly smaller 
than unity.

\section{Discussion} \label{discussion}
SNR \snr\ has been found to be a $\gamma$-ray emitting source in the band
from GeV to sub-PeV
\citep[e.g.,][]{xin19, alb20, ame21, cao21}. The main part of the $\gamma$-ray
emission appears to be spatially correspondent with 
the CO emission in LSR velocity 
interval $-6$ to $-3\km\ps$
projected
in the ``belly'' region of
the SNR \citep[e.g.,][]{xin19, ame21}. 
By parameterizing the distance to the MC as $\du = d /$(0.8\,kpc)
\citep{kot01}, the mass of the molecular gas in velocity range $-9$ 
to $-1\km\ps$ within the cyan rectangle region delineated in
Figure~\ref{ratio} is estimated to be about 
$330\du M_\odot$ and $860\du M_\odot$ from \thCO\ and \twCO\ emissions
\footnote{The mass of the MC estimated using \twCO\ here could have
large uncertainty, as the X-factor method is more reliable for MC with
a mass of order larger than $10^4M_\odot$ \citep{bol13}.}
respectively (see Appendix~\ref{mass}).
Assuming the length of the cloud along the line of sight (LOS) is comparable 
to its projected size ($\sim 4.8\pc$), the molecular density of the cloud 
is estimated to be about $30\pcc$ and $80\pcc$, respectively.
Our observation also
shows that a filament follows the northern radio 
boundary of the SNR in velocity interval $-8$ to $-5\km\ps$. 
Asymmetrical \twCO\ line profiles 
are found in some locations along the northern boundary and in the
``belly'' region of \snr, but we could not conclude that they represent
the kinematic evidence of shock-MC interaction since we also see similar
\twCO\ line profiles outside the SNR (Figure~\ref{regsed}).
Physical signature of interaction 
(such as elevated gas temperature or \twCO\ \tto/\otz\ line ratio)
has not been found either along the
filament or in the MC toward the ``belly'' of the SNR
(Figures~\ref{tkin} and \ref{ratio}).
Nonetheless, we can not exclude the possible proximity of the MC to the SNR 
along the LOS. 
In such a scenario, the accelerated protons that escape from the SNR 
shock can illuminate the MC via p-p hadronic interaction.

Except for SNR(s), other possible Galactic CR accelerators, such as
cluster(s) with massive stars \citep[e.g.,][]{aha19, mor21}, are also 
worth considering.
We note that there are structural properties in the MCs toward the 
gamma-ray emitting region in the velocity range of about $-11$ to $-3\km\ps$ 
(see \S~\ref{spatial} and \ref{pvresult}). 
Especially, an expanding molecular structure and a velocity
gradient across the ``belly'' region of \snr\ in
the P-V diagram (\S~\ref{pvresult})
might be the results of massive stellar winds.
The radius $R$ and the expansion velocity $v$ 
of the expanding sturcture are about $12'$ (corresponding 
to $\sim 2.9\du$\,pc) and $3.5\km\ps$, respectively (\S~\ref{pvresult}). 
If it is blown by the interior stellar wind(s), 
its age could be estimated to be \citep{wea77}
$t = 3R/5v = 5\times 10^5 \du (v/3.5\km\ps)^{-1} \yr$. 
Furthermore, the kinetic luminosity of the stellar wind(s) could be 
\citep{wea77}:
$L_w \sim 1.2\times 10^{34} (n/100\pcc) (v/3.5\km\ps)^3 \du ^2 \erg\ps$,
where $n$ is the atomic hydrogen number density of the molecular gas.
The estimated luminosity of the stellar wind(s) is typical for 
an O9V to B0V type star \citep[e.g.,][]{els89,che13}.
We have searched for massive star(s) using the \emph{Tycho}-2 Spectral Type 
Catalog \citep{wri03} toward this region, yet no OB star cluster/association 
or massive main sequence star has been found,
and it is unlikely to be a result of incompleteness of the catalog we 
use (see Appendix~\ref{obstar}).
Therefore, we suggest that the expanding structure
and the velocity gradient of the molecular gas could be
due to the progenitor wind of \snr.
We note that similar structures due to progenitor
winds have also been found in a few other SNRs, such as Tycho
\citep{zho16, che17}, N132D \citep{san20}, and VRO 42.05.01 \citep{ari19}.

\section{Summary} \label{summary} 
We have investigated the molecular environment of \snr\ using the CO-line
data obtained with the IRAM 30\,m telescope toward 
the ``belly'' region of the SNR, which is coincident with the 
centroid region of the $\gamma$-ray emission.
There is a filament that follows the north
radio boundary of the SNR in velocity interval $-8$ to $-5\km\ps$.
Some asymmetric broad profiles of \twCO\ lines
are found not only toward the ``belly'' region of 
the SNR but also outside the boundary. 
Further, the \twCO\ \tto/\otz\ line ratios of the MCs are similar
to the typical value for interstellar MCs. 
Therefore, it is uncertain about whether the MCs are disturbed 
by the SNR shock.
Nonetheless, we can not exclude the 
possible proximity of the MC to the SNR along the LOS, which 
facilitates illumination of the MC by the accelerated
protons that escape from the SNR shock via p-p hadronic interaction.
Notably, we find an expanding molecular structure 
and a velocity gradient toward the 
``belly'' region of the SNR, and we
propose a possibility of their relations
to the progenitor wind of the SNR.

\begin{acknowledgements}
We are grateful to Xiao Zhang for the discussion about the gamma-ray 
emission toward this SNR, Xin Zhou for the discussion about the methods
to estimate the properties of the MCs,
Niu Liu and Yong Shao for a discussion about how to
search for OB stars in the field of view. 
We thank the staff of the IRAM 30\,m observatory for the help during the 
remote observation.
This work is supported by the National Key R\&D
Program of China under grants 2017YFA0402600, and the NSFC under
grants 11773014, 11633007, 11503008, and 11590781.
Q.C.L. acknowledges
support from the program A for Outstanding PhD candidate of Nanjing University.
\end{acknowledgements}


%
%
%

\bibliographystyle{aasjournal}

\bibliography{ref}

\appendix
\renewcommand\thefigure{\thesection.\arabic{figure}}    
\setcounter{figure}{0}

\section{Grid of the CO spectra observed by the IRAM 30\,m telescope}
\label{linegrid}
\begin{figure}
\centering
\includegraphics[scale=.3]{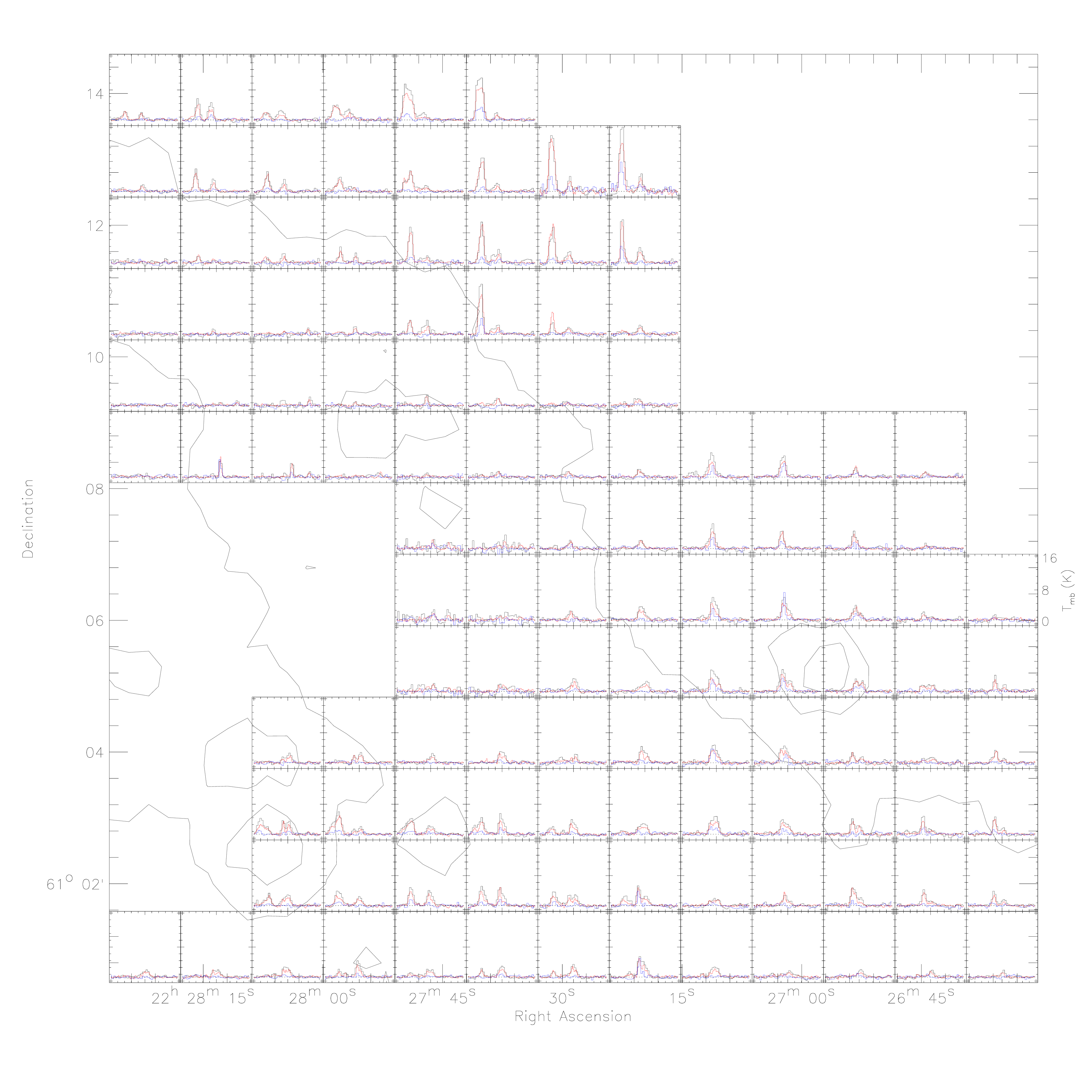}
\caption{
Grid of \twCO\ \otz, \twCO\ \tto, and \thCO\ \otz\ spectra observed
by the IRAM 30\,m telescope in velocity range $-15$ to $+5\km\ps$ for the
region delineated by the white rectangle labeled with ``N'' in the right
panel of Figure~\ref{tkin}. The black lines denote the \twCO\ \otz\ spectra,
the red lines denote the \twCO\ \tto\ spectra, and the blue lines denote
the \thCO\ \otz\ spectra. The \thCO\ \otz\ spectra have been multiplied by
a factor of two for better visibility. The size of each pixel is $66''\times 66''$.
The contours are the same as those in Figure~\ref{multi}.
}
\label{linegrid_N}
\end{figure}

\begin{figure}
\centering
\includegraphics[scale=.3]{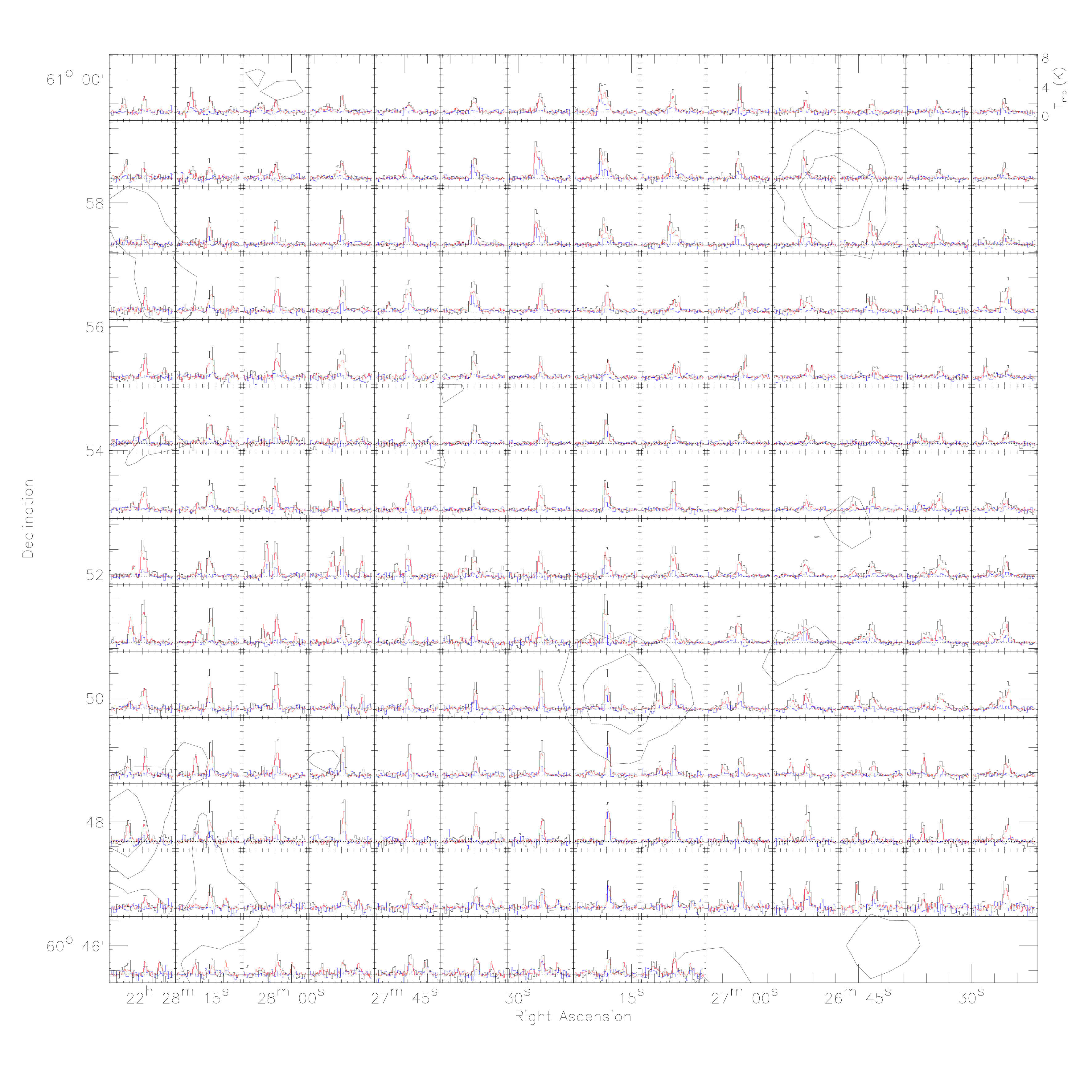}
\caption{
The same as Figure~\ref{linegrid_N}, but for the region delineated by the
white rectangle labeled with ``S'' in the right panel of Figure~\ref{tkin}.
The \thCO\ \otz\ spectra also have been multiplied by a factor of two for 
better visibility.
}
\label{linegrid_S}
\end{figure}

\begin{figure}
\centering
\includegraphics[scale=.3]{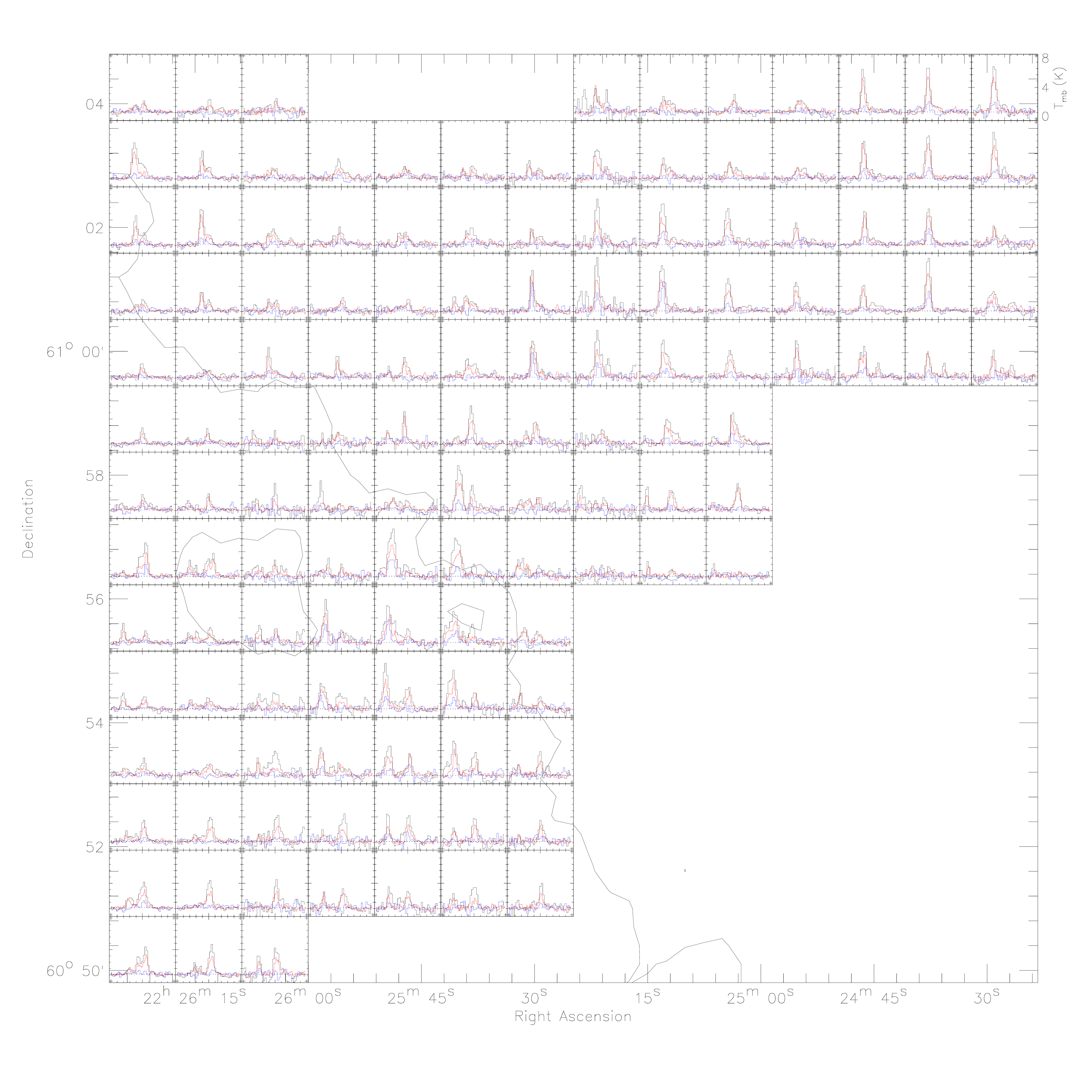}
\caption{
The same as Figure~\ref{linegrid_N}, but for the region delineated by the
white rectangle labeled with ``W'' in the right panel of Figure~\ref{tkin}.
The \thCO\ \otz\ spectra also have been multiplied by a factor of two for 
better visibility.
}
\label{linegrid_W}
\end{figure}

We have inspected the CO line profiles toward the SNR \snr\ region
observed by the IRAM 30\,m telescope. Because the pixel sizes of the 
original datacubes are too small ($11''$), we have re-gridded the datacubes
to reach pixel sizes of $66''$ before making the grids of \twCO\ \otz,
\twCO\ \tto, and \thCO\ \otz\ spectra. Figures~\ref{linegrid_N}, \ref{linegrid_S},
and \ref{linegrid_W} show the grids of CO spectra in velocity range $-15$
to $+5\km\ps$ for the regions delineated by the white rectangles labeled with
``N'', ``S'', and ``W'' in the right panel of Figure~\ref{ratio}, respectively.
There are many pixels, either within, on, or outside the boundary of the SNR
\snr, where the profiles of \twCO\ spectra seemingly to be asymmetric broadened.
The averaged CO line profiles of some of the pixels (regions ``1'' to ``8'' marked
in Figures~\ref{tkin} and \ref{ratio}) are studied in detail in 
Section~\ref{line}.

\section{Temperature Distribution of the Molecular Clouds}
\label{temperature}
\begin{figure}
\centering
\includegraphics[scale=.9]{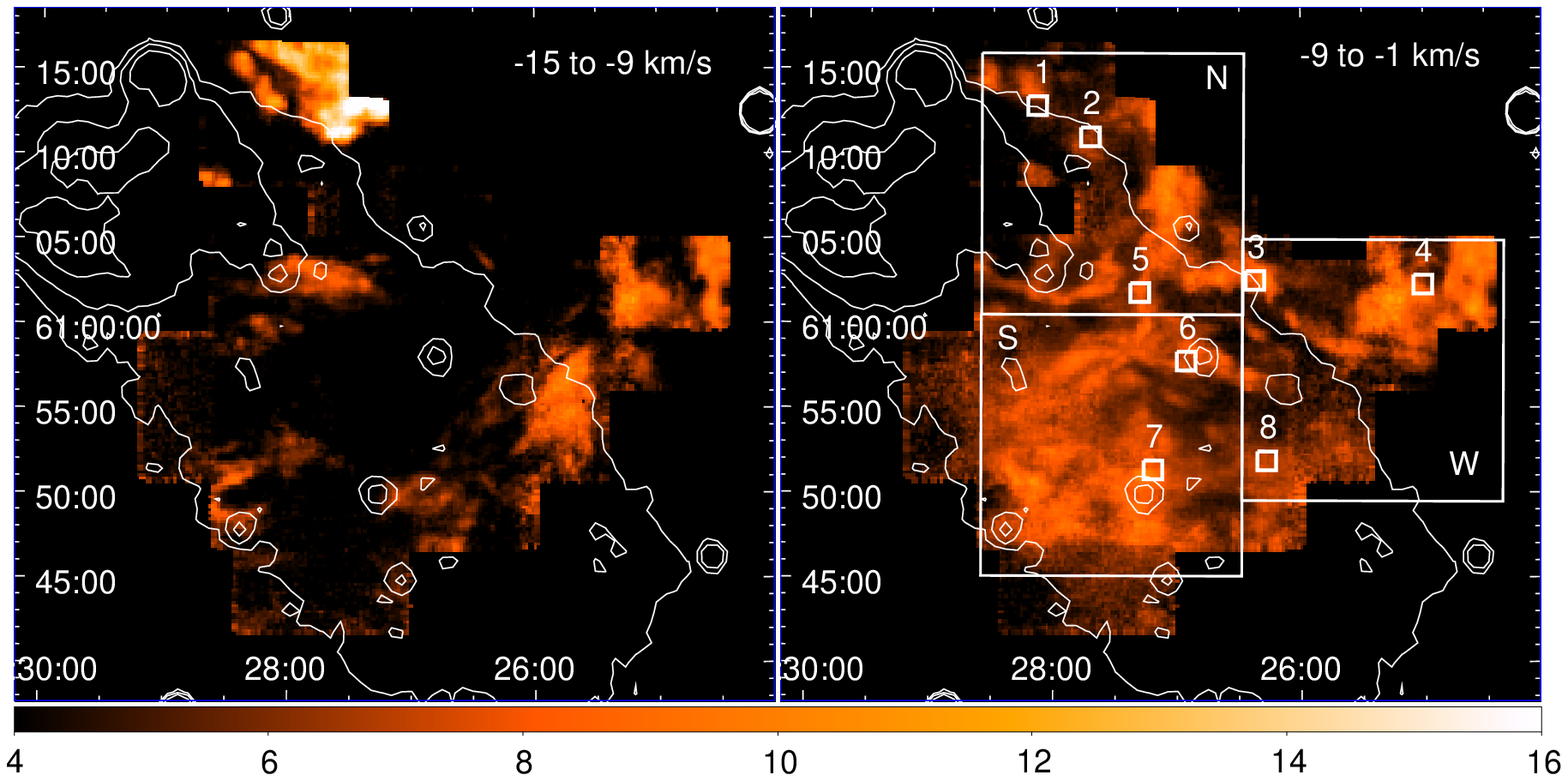}
\caption{ Kinetic temperature ($T_{\rm k}$) maps of the MCs in velocity intervals 
$-15$ to $-9\km\ps$ (left panel) and $-9$ to $-1\km\ps$ (right panel) obtained from 
the IRAM 30\,m telescope observation, with the assumptions of no beam dilution ($f=1$) and
$\tau_{^{12}CO} \gg 1$. 
The contours are the same as in Figure~\ref{multi}.
The three regions delineated by large white rectangles and labeled with ``N'',
``S'', and ``W'' are used to extract the CO grid spectra (Figure~\ref{linegrid_N},
\ref{linegrid_S}, and \ref{linegrid_W}). The regions delineated by
small white
rectangles are the same as those in Figure~\ref{ratio}.
}
\label{tkin}
\end{figure}

The kinetic temperature of the molecular gas ($T_{\rm k}$), which is 
approximately equal to the excitation temperature ($T_{\rm ex}$) under 
the assumption of local thermodynamic equilibrium, 
can be estimated from the peak of the main-beam 
temperature $T_{\rm mb}$ of the \twCO\ \otz\ line emission with equation
$T_{\rm k} = 5.53/\{\ln [1+5.53/(T_{\rm mb}+0.84)]\}\,{\rm K}$,
where the beam filling factor is assumed to be equal to 1. 
Figure~\ref{tkin} shows the spatial distribution of the kinetic temperature 
of the molecular gas in velocity intervals $-15$ to $-9\km\ps$ 
and $-9$ to $-1\km\ps$, corresponding to the velocity ranges of the two 
components shown in the top left panel of Figure~\ref{chanelmap}. 
The kinetic temperature in each pixel is estimated from
the peak of the main-beam temperature in the given velocity range.
The kinetic temperatures are $\la 10$\,K for the most part 
of the molecular gas in both velocity ranges, except for the clump at the 
northern corner in velocity interval $-15$ to $-9\km\ps$.
That is, the distribution of the kinetic temperature
is similar to that of the \twCO\ \tto/\otz\ line ratios.
Actually, the higher line ratio toward the northern corner
in this velocity interval could be directly related to the higher
kinetic temperature there (see, e.g., \citealt{liu20}).

\section{The Mass of the Molecular Cloud toward the ``belly'' of \snr}
\label{mass}
The mass of the molecular gas in velocity range $-9$ to $-1\km\ps$
toward the ``belly'' of \snr\ could be 
roughly estimated from the column density of the MC in the given region.
The angular
size of the region delineated by the cyan rectangle in the right panel
of Figure~\ref{ratio} is $\sim 18'.5\times 22'.5$, which can be translated
to a physical size of $\sim 4.3\du\pc\times 5.2\du\pc$. The column density
of H$_2$ within this region is estimated from either 
\thCO\ \otz\ or \twCO\ \otz\ line. When assuming the molecular gas to be
in local thermodynamic equilibrium and the \twCO\ \otz\ line to be optically
thick, the column density could be estimated using 
$N\HH\approx 7\times 10^5 N(^{13}{\rm CO})$ \citep{fre82}.
Here $N(^{13}{\rm CO}) = 2.42\times 10^{14} \int T_{\rm mb}
(^{13}{\rm CO}) dv/(1-{\rm exp}(-5.3/T_{\rm ex}))\psc$
is the column density of $^{13}$CO.
On the other hand, it could also be estimated using the CO-to-H$_2$ mass
conversion factor ($N\HH/W(^{12}{\rm CO})$; the ``X-factor''), which 
is assumed to be $1.8\times 10^{20}\psc {\rm K}^{-1}\km ^{-1}\,{\rm s}$
\citep{dam01}. The mass of this MC is thus estimated to be 
about $330\du M_\odot$ and $860\du M_\odot$ from \thCO\ and \twCO\
lines, respectively.

\section{The Massive Stars Toward \snr} \label{obstar}
We use the \emph{Tycho}-2 Spectral Type Catalog \citep{wri03}, which is 
about 90\% complete at an apparent visual magnitude of $\sim 11.5$ 
\citep{hog00}, to search for massive stars that can potentially 
produce wind-blown structure (O--B3V star; e.g., \citealt{che13})
toward \snr\ in a circular region centered at 
R.A.=22\h27\m12\s.58, decl.=+60\ci46$'19.9''$, $30'$ in radius.
However, no massive star has been
found toward \snr, which is unlikely to be a result of optical extinction.
The optical extinction can be estimated from the empirical relation
$N_{\rm H}=(2.87\pm 0.12) \times 10^{21}A_V \psc$ \citep{foi16},
where $N_{\rm H}=2\NHH +N_{\rm HI}$ 
($\sim 9.5\times 10^{21}\psc$, \citealt{ge20}) is the
foreground hydrogen 
(including atoms and molecules) column density. 
Therefore, the optical extinction can be
estimated to be 
$A_V=3.31\pm 0.15$.
According to the relation $V=M_V-5+5\log D+A_V$, where $V$ is the apparent
visual magnitude, $M_V$ is the absolute magnitude 
($\le -1.68$ for stars earlier than B3V type;
e.g., \citealt{weg06}), $D$ is the distance to the star in units of pc,
the apparent visual magnitude of a massive star at the distance of \snr\
(about 800\,pc; \citealt{kot01}) $\le 11.1$,
which should have been smaller than 11.5.

\end{document}